\documentclass[12pt]{iopart}
\usepackage{iopams}
\usepackage{graphicx}
\usepackage{epsfig}

\begin{document}

\title{Spinons and Holons with Polarized Photons in a Nonlinear Waveguide}
\author{ Ming-Xia Huo$^{[1]}$, Dimitris G. Angelakis$^{[1,2]}$, and Leong
Chuan Kwek$^{[1,3]}$}

\address{$^{1}$ Centre for Quantum Technologies, National University of
Singapore, 3 Science Drive 2, Singapore 117542}
\address{$^{2}$ Science
Department, Technical University of Crete, Chania, Crete, Greece,}
\address{$^{3}$ National Institute of Education and Institute of Advanced Studies, Nanyang
Technological University, 1 Nanyang Walk, Singapore 637616}

\footnote{%
contact: dimitris.angelakis@gmail.org}



\begin{abstract}
We show that the spin-charge separation predicted for correlated fermions in
one dimension, could be observed using polarized photons propagating in a
nonlinear optical waveguide. Using coherent control techniques and employing
a cold atom ensemble interacting with the photons, large nonlinearities in
the single photon level can be achieved. We show that the latter can allow
for the simulation of a strongly interacting gas, which is made of
stationary dark-state polaritons of two species and then shown to form a
Luttinger liquid of effective fermions for the right regime of interactions.
The system can be tuned optically to the relevant regime where the
spin-charge separation is expected to occur. The characteristic features of
the separation as demonstrated in the different spin and charge densities
and velocities can be efficiently detected via optical measurements of the
emitted photons with current optical technologies.
\end{abstract}

\maketitle

\section{Introduction}

\subsection{Spin-charge separation and quantum simulators}

One-dimensional (1D) physical systems have attracted much attention due to
their novel and sometimes spectacular features. Unlike two- or
three-dimensional systems, the physics of 1D Lieb-Liniger model \cite{Lieb}
is well captured by the Luttinger liquid theory in the low-energy domain
\cite{Girardeau1}. In 1D Luttinger liquid theory, collective excitations
rather than single excitations appear due to the tight transverse
confinement forcing particles to move along one direction and thus
converting any individual motion into a collective one. The collective modes
of spin and charge in the 1D electronic gases surprisingly can be shown to
propagate with different velocities, \textit{i.e.}, the spin and the charge
separate into spinons and holons \cite%
{Girardeau1,Girardeau2,Girardeau3,Girardeau4}. In experiments, the
observation of spin-charge separation is however challenging - the control
of interactions is still elusive and no distinct features of separation are
obtained, although several seminal works have been made using copper-oxide
chain compounds SrCuO$_{2}$ \cite{SC_exp_old_77}, metallic chains \cite%
{SC_exp_old_402}, superconductors \cite{SC_exp_old_418}, and more recently,
GaAs/AlGaAs heterostructures \cite{SCC4_308,SCC4_2,SCC4_325}.

At the same time, works on artificially engineered quantum optical systems
in which many-body effects could be reproduced in well controlled
environments have recently emerged. To observe spin-charge separation in
cold atoms, experimental proposals involving bosonic and fermionic species
have been suggested \cite{Recati1,Recati2,Recati3,Recati4}. However, the
challenges in the trapping and cooling of fermionic gases, especially the
lack of necessary individual accessibility and measurement of correlations
in general, make current results inconclusive. Strongly correlated photons
and polaritons, as hybrid light-matter quantum simulators, have been
recently proposed. Initially using coupled resonator implementations \cite%
{SIPS1}, Mott transitions \cite{SIPS21,SIPS22,SIPS23} and then effective
spin models and Fractional Hall states of light were shown to be possible
\cite{SIPSother1,SIPSother2,SIPSother3,SIPSother4,SIPSother5,SIPSother6}.
More recently, by employing hollow-core optical waveguides filled with cold
atom ensembles and using slow-light techiques \cite{DPS1,DPS2,DPS3}, it was
shown that it is possible to prepare a Tonks gas of photons \cite{Chang}.
Using two atomic species, a two-component Lieb-Liniger model has also been
recently suggested \cite{SCS}. Compared to cold-atom proposals, photonic
proposals should allow for more direct measurements of local observables and
correlation functions of the emitted photon states.

In this article, we consider a novel scheme involving two oppositely
circularly polarized quantum beams and single atomic species to simulate a
two-component interacting gas in 1D. We note the difference to an earlier
scheme, where two quantum fields with different frequencies interacting with
two species of four-level atoms were employed \cite{SCS}. The current
proposal has the distinct advantage over in two major ways. Firstly in terms
of more efficient detection of the correlation in output polarized states
and secondly in terms of the loading and preparation process into the fiber
which requires handling a single atomic species rather than two.

The paper is organized as follows. We first review the basics of the single-
and two-component bosonic Lieb-Liniger model \cite{Lieb}, and especially its
mapping to Luttinger liquid theory. We then describe in detail the
preparation of a polaritonic Lieb-Liniger model with two bosonic components
in the waveguide employing slow-light techniques. Finally we discuss the
necessary range of optical parameters in order to drive the system to the
relevant spin-charge separation regime. We conclude by analyzing how the
effective photonic spin and charge densities and velocities can be probed.
This is done by releasing the trapped polaritons into outgoing photons where
optical measurements can reconstruct the characteristic functions of the
effect.

\subsection{ From Lieb-Liniger bosons to Luttinger liquids: The basics of
Spin-Charge Separation revisited}

One of the most famous 1D models is the Lieb-Liniger model \cite{Lieb},
which describes $N$ bosons with a Dirac-delta interaction as
\begin{equation}
H^{\mathrm{s}}=-\frac{1}{2m}\sum_{i=1}^{N}\frac{\partial ^{2}}{\partial
z_{i}^{2}}+g\sum_{i<j=1}^{N}\delta \left( z_{i}-z_{j}\right) ,
\end{equation}%
where $m$\ is the mass and $g$\ is the interaction strength. Although the
Lieb-Liniger model is exactly solvable through the Bethe ansatz approach, it
is still generally very difficult to extract correlation functions from the
solutions at any interaction regime. Luttinger liquid theory on the other
hand can give the low energy universal description of the Lieb-Liniger model
at low temperatures. In the Luttinger liquid phase, the low energy
excitations are no longer single but collective modes with linear
dispersion. The confinement of interactions between particles in 1D forces
any individual motion to affect the collective system. The description of
the dynamics in terms of collective bosonic fields is called `bosonization'
approach and we briefly present it in the following.

Assume $N$ particles described by $\psi (\mathbf{r})=\psi (z)\varphi _{0}(%
\mathbf{r}_{\perp })$, where the particles move along one direction, say $z$%
\ direction. Strong confinement is applied to the transverse direction $%
\mathbf{r}_{\perp }=\{x,y\}$, allowing only for the lowest energy transverse
quantum state $\varphi _{0}(\mathbf{r}_{\perp })$\ to be considered. The
Lieb-Liniger model of the parallel component reads \cite%
{Girardeau1,Girardeau2,Girardeau3,Girardeau4}
\begin{equation}
H^{\mathrm{s}}=\int dz[\frac{1}{2m}\partial _{z}\psi ^{\dagger }\left(
z\right) \partial _{z}\psi \left( z\right) +\frac{g}{2}\rho ^{2}\left(
z\right) ],  \label{LLP}
\end{equation}%
where the collective bosonic fields $\psi \left( z\right) \ $and $\psi
^{\dagger }\left( z\right) $ can be expressed as%
\begin{equation}
\psi \left( z\right) =e^{-i\theta \left( z\right) }\left[ \rho \left(
z\right) \right] ^{1/2},\psi ^{\dagger }\left( z\right) =\left[ \rho \left(
z\right) \right] ^{1/2}e^{i\theta \left( z\right) },
\end{equation}%
with $\rho \left( z\right) =\psi ^{\dagger }\psi $ the particle density and $%
\theta \left( z\right) $ the phase operator. As described in \cite%
{Girardeau1}, the phase and density fields are canonically conjugated fields,%
\begin{equation}
\lbrack \rho (z),e^{i\theta (z^{\prime })}]=\delta (z-z^{\prime })e^{i\theta
(z^{\prime })},  \label{cg}
\end{equation}%
and the collective density operator can be expressed as
\begin{equation}
\rho (z)=[\rho _{0}+\frac{1}{\pi }\partial _{z}\phi (z)]\sum_{m=-\infty
}^{+\infty }\exp \{im[2\pi \rho _{0}z+2\phi (z)]\}.  \label{do}
\end{equation}%
If $\rho (z)$\ varies slowly with z, we can retain the lowest frequency
component for $m=0$ and write
\begin{equation}
\rho (z)\simeq \rho _{0}+\frac{1}{\pi }\partial _{z}\phi (z),  \label{do2}
\end{equation}%
where the fields $\theta (z)$ and $\frac{1}{\pi }\partial _{z}\phi
(z^{\prime })$\ are canonically conjugated.

The Lieb-Liniger Hamiltonian (\ref{LLP}) for the lowest component $\psi
(z)\simeq e^{-i\theta \left( z\right) }\rho _{0}^{1/2}$ can be mapped to a
Luttinger liquid with Hamiltonian (see \cite%
{Girardeau1,Girardeau2,Girardeau3,Girardeau4}):
\begin{equation}
H^{\mathrm{s}}=\int \frac{dz}{2\pi }[\upsilon K^{\mathrm{sl}}(\partial
_{z}\theta )^{2}+\frac{\upsilon }{K^{\mathrm{sl}}}(\partial _{z}\phi )^{2}],
\end{equation}%
where all the interaction effects are encoded into two effective parameters:
the propagation velocity of density disturbances $\upsilon $\ and the
so-called Luttinger parameter $K^{\mathrm{sl}}$ controlling the
long-distance decay of correlations.

Going beyond the simple case with single-component bosonic system,
interesting physics can be obtained by mixing two bosonic components or by
considering two internal degrees of freedom of bosons, which is analogous to
assigning a \textquotedblleft spin\textquotedblright $1/2$\ to bosons. The
two-component Lieb-Liniger model in this case would read:%
\begin{equation}
H^{\mathrm{t}}=\int dz\sum_{s=\uparrow ,\downarrow }\left[ \frac{1}{2m_{s}}%
\partial _{z}\psi _{s}^{\dagger }(z)\partial _{z}\psi _{s}(z)+\frac{\chi _{s}%
}{2}\rho _{s}^{2}(z)\right] +\int dz\chi _{\uparrow \downarrow }\rho
_{\uparrow }(z)\rho _{\downarrow }(z),
\end{equation}%
where $m_{s}$ is the mass and $\chi _{s}$\ and $\chi _{\uparrow \downarrow }$%
\ are the intra- and interspecies interaction with $s=\uparrow $, $%
\downarrow $\ representing the two spins. Following the literature \cite%
{Girardeau1,Girardeau2,Girardeau3,Girardeau4}, the charge- and spin-related
fields can be defined as:
\begin{equation}
\theta _{\mathrm{charge}}=\frac{\theta _{\uparrow }+\theta _{\downarrow }}{%
\sqrt{2}},\theta _{\mathrm{spin}}=\frac{\theta _{\uparrow }-\theta
_{\downarrow }}{\sqrt{2}},\phi _{\mathrm{charge}}=\frac{\phi _{\uparrow
}+\phi _{\downarrow }}{\sqrt{2}},\phi _{\mathrm{spin}}=\frac{\phi _{\uparrow
}-\phi _{\downarrow }}{\sqrt{2}}.
\end{equation}%
The Hamiltonian separates into two parts (more details in \cite{Girardeau1})
and reads as: $H^{\mathrm{t}}=H_{\mathrm{charge}}+H_{\mathrm{spin}}$. The
charge part is given by
\begin{equation}
H_{\mathrm{charge}}=\int \frac{dz}{2\pi }[u_{\mathrm{charge}}K_{\mathrm{%
charge}}(\partial _{x}\theta _{\mathrm{charge}})^{2}+\frac{u_{\mathrm{charge}%
}}{K_{\mathrm{charge}}}(\partial _{x}\phi _{\mathrm{charge}})^{2}]
\end{equation}%
and the spin part is defined as%
\begin{eqnarray}
H_{\mathrm{spin}} &=&\int \frac{dz}{2\pi }[u_{\mathrm{spin}}K_{\mathrm{spin}%
}(\partial _{x}\theta _{\mathrm{spin}})^{2}+\frac{u_{\mathrm{spin}}}{K_{%
\mathrm{spin}}}(\partial _{x}\phi _{\mathrm{spin}})^{2}]  \nonumber \\
&&+2\chi _{\uparrow \downarrow }\rho _{0}^{2}\cos \left( \sqrt{8}\phi _{%
\mathrm{spin}}\right)
\end{eqnarray}%
under the separation conditions
\begin{equation}
\chi _{\uparrow }=\chi _{\downarrow },\frac{\rho _{0,\uparrow }}{m_{\uparrow
}}=\frac{\rho _{0,\downarrow }}{m_{\downarrow }}.  \label{con}
\end{equation}%
Here $\rho _{0}=\rho _{0,\uparrow }+\rho _{0,\downarrow }$ and $\rho _{0,s}$
are the densities for two species. With $\chi =\chi _{s}$, $u=u_{s}=\sqrt{%
\rho _{0,s}\chi _{s}/m_{s}}$, and $K=K_{s}=\sqrt{\pi ^{2}\rho
_{0,s}/(m_{s}\chi _{s})}$, the charge and spin velocities and Luttinger
parameters are $u_{\mathrm{charge,spin}}=u\sqrt{1\pm \chi _{\uparrow
\downarrow }/\chi },$ $K_{\mathrm{charge,spin}}=K/\sqrt{1\pm \chi _{\uparrow
\downarrow }/\chi }$.

%
%


\section{ Photonic Spin-Charge Separation in Nonlinear Optical Waveguides}

\subsection{The Optical Waveguide Setup}

Our proposal is based on exploiting the available huge photonic
nonlinearities possible to generate in specific quantum optical setups. More
specifically, we envisage the use of a highly nonlinear waveguide where the
necessary nonlinearity will emerge through the strong interaction of the
propagating photons to existing emitters in the waveguide. Recent
experiments have developed two similar setups in this direction, both
capable of implementing our proposal with either current or near future
platforms. In these experiments, cold atomic ensembles are brought close to
the surface of a tapered fiber \cite{Nayak,Vetsch} or are loaded inside the
core of a hollow-core waveguide \cite%
{book_Menzel1,book_Menzel3,book_Menzel4,book_Menzel5} as shown in figure \ref%
{al}\ (a). The available optical nonlinearity based on the
Electromagnetically Induced Transparency (EIT) effect can be used as we will
show to create situations where the trapped photons obey Lieb-Liniger
physics.


The process to generate the strongly correlated states of photons is as
follows: First, laser-cooled atoms exhibiting a multiple atomic-level
structure shown in figure \ref{al}\ (b) are moved into position so they will
interact strongly with incident quantum light fields. Initially, resonant to
the corresponding transitions, two optical pulses with opposite
polarizations $\hat{E}_{\uparrow ,+}(z,t)$ and $\hat{E}_{\downarrow ,+}(z,t)$%
\ are sent in from one direction, say the left side. They are injected into
the waveguide with the co-propagating classical control fields $\Omega
_{\uparrow ,+}(t)$ and $\Omega _{\downarrow ,+}(t)$ initially turned on. As
soon as the two quantum pulses completely enter into the waveguide, the
classical fields $\Omega _{s,+}$ are adiabatically turned off, converting $%
\hat{E}_{s,+}$\ into coherent atomic excitations as in usual slow-light
experiments for $s=\uparrow ,\downarrow $. We then adiabatically switch on
both $\Omega _{s,+}$\ and $\Omega _{s,-}$\ from two sides. The probe pulses
become trapped due to the effective Bragg scattering from the stationary
classical waves as analyzed in \cite{DPS1,DPS2,DPS3}. At this stage the
pulses are noninteracting with the photons expanding freely due to the
dispersion. By slowly shifting the $d$-levels, the effective masses can be
kept constant whereas the effective intra- and interspecies repulsions are
increased. This drives the system into a strongly interacting regime. This
dynamic evolution is possible by keeping for example the corresponding
photon detunings $\Delta _{s}$\ constant while shifting the $d$-level. Once
this correlated state is achieved, the fields - for example $\Omega _{s,+}$
- from the pair of control fields that trap polaritons, are slowly turned
off. This will release the corresponding quasi-particles by turning them to
propagating photons which will then exit the fiber. As all correlations
established in the previous step are retained, these wave packets comprise
of two separated effective charge and spin density waves.

\subsection{Realizing a two-component Lieb-Liniger model of polarized photons%
}

\begin{figure}[tbp]
\centering
\includegraphics[width = 0.75\linewidth]{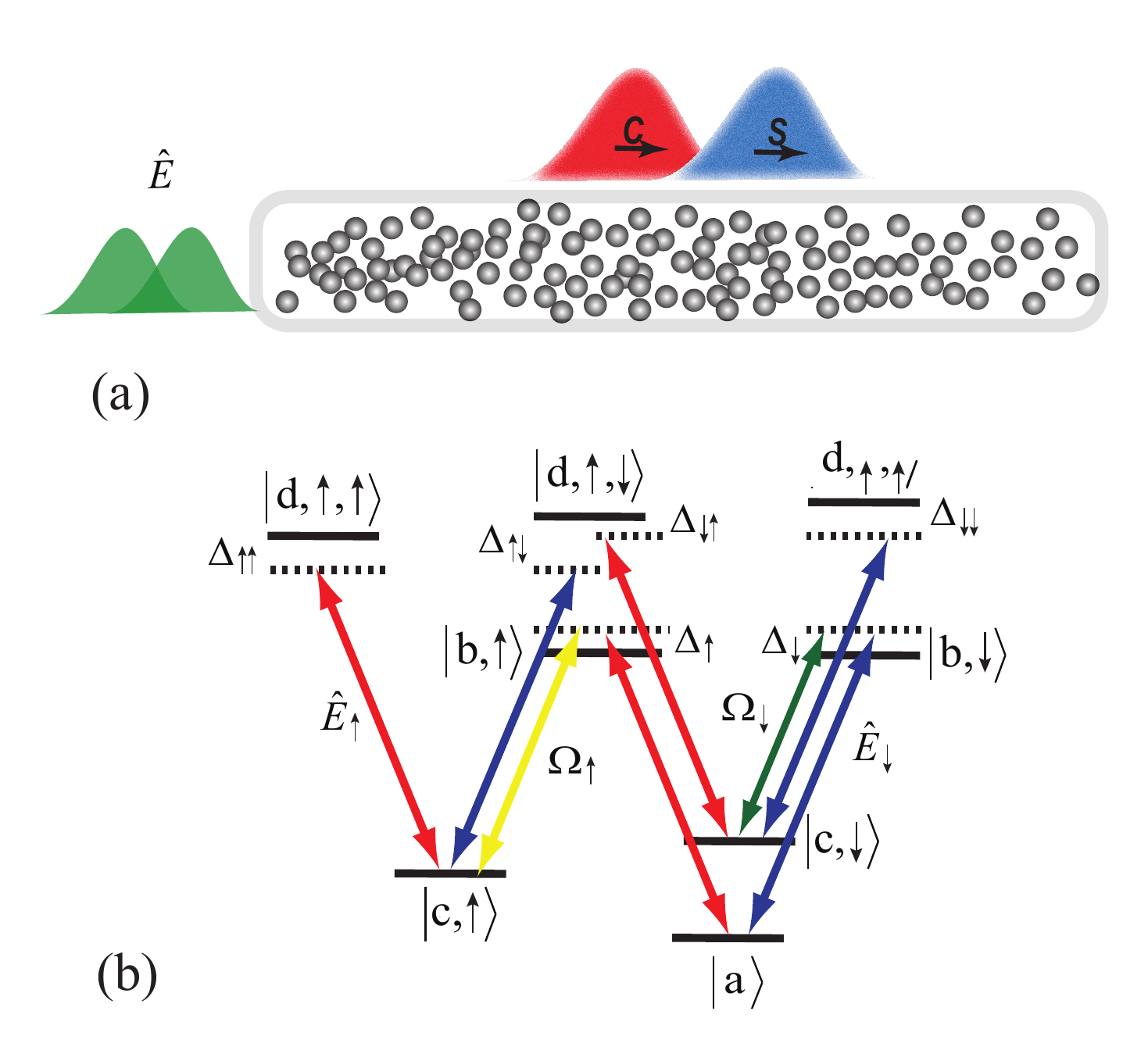}
\caption{The model setup under consideration. In a nonlinear fiber ( a
hollow core version is shown here but a tapered fiber approach could also be
used), the cold atoms are interacting with two quantum light fields $\hat{E}%
_{s}$ (red and blue arrowlines) and two pairs of classical fields $\Omega
_{s,\pm }$ (yellow and green arrowlines), where $\pm \ $denotes the forward
or backward propagation direction. The studied atomic level diagram and
possible atomic transitions driven by two oppositely circularly polarized
quantum pulses $\hat{E}_{s}$\ and two control beams $\Omega _{s}$ is shown
in (b). Appropriate tuning the couplings of light fields to the
corresponding atomic transitions, forces the trapped polaritons to behave as
an effective 1D quantum Luttinger liquid and reach the spin-charge
separation regime. Coherently transferring the polaritons' correlations to
propagating light pulses and allowing them to exit the fiber, provide for
the efficient measurement of both the dynamics of the propagation of the
effective spin and charge quasiparticles and the effective spin and charge
velocities characteristic of the effect taking place.}
\label{al}
\end{figure}

The system described above and shown in figure \ref{al} obeys the
Hamiltonian:%
\begin{eqnarray}
H^{\mathrm{o}} &=&-\int \sum_{s}n_{z}^{s}dz\{-\Delta _{s}\sigma
_{b,s;b,s}-\sum_{s^{\prime }}\Delta _{ss^{\prime }}\sigma _{d,s,s^{\prime
};d,s,s^{\prime }}^{\mathnormal{x}}  \nonumber \\
&&+\sqrt{2\pi }(g_{s}\sigma _{b,s;a}+\sum_{s^{\prime }}g_{ss^{\prime
}}\sigma _{d,s,s^{\prime };c,s^{\prime }})  \nonumber \\
&&\times \left( \hat{E}_{s,+}\mathrm{e}^{\mathrm{i}k_{\mathrm{Q}\mathnormal{%
,s}}z}+\hat{E}_{s,-}\mathrm{e}^{-\mathrm{i}k_{\mathrm{Q}\mathnormal{,s}%
}z}\right)  \nonumber \\
&&+\left( \Omega _{s,+}\mathrm{e}^{\mathrm{i}k_{\mathrm{C}\mathnormal{,s}%
}z}+\Omega _{s,-}\mathrm{e}^{-\mathrm{i}k_{\mathrm{C}\mathnormal{,s}%
}z}\right) \sigma _{c,s;b,s}+\mathrm{h.c.}\},
\end{eqnarray}%
where $s,s^{\prime }=\uparrow ,\downarrow $ and $\left\vert d,\uparrow
,\downarrow \right\rangle =\left\vert d,\downarrow ,\uparrow \right\rangle $%
. The continuous collective atomic spin operators $\sigma _{\mu ;\nu }\equiv
\sigma _{\mu ;\nu }(z,t)$ describe the averages of the flip operators $%
\left\vert \mu \right\rangle \left\langle \nu \right\vert $ over atoms in a
small region around $z$. The density of atoms is $n_{z}^{\mathnormal{s}}$
and $g_{s}$, $g_{ss^{\prime }}$ are the coupling strengths between the
quantum fields and atoms, while $\Delta _{s}$ and $\Delta _{ss^{\prime }}$
are one-photon detunings from the corresponding transitions. For simplicity,
we assume that $g_{s}=g_{ss^{\prime }}=g$. Furthermore, we label the two
quantum and two classical fields with frequencies $\omega _{\mathrm{Q}%
\mathnormal{,s}}$ and $\omega _{\mathrm{C}\mathnormal{,s}}$ and wave vectors
$k_{\mathrm{Q}\mathnormal{,s}}$ and $k_{\mathrm{C}\mathnormal{,s}}$,
respectively. Both quantum fields $\hat{E}_{\uparrow ,+}(z,t)$ and $\hat{E}%
_{\downarrow ,+}(z,t)$ drive four possible atomic transitions. The fields $%
\hat{E}_{s,\pm }(z,t)$ are detuned by $\Delta _{s}$ from the transition $%
|a\rangle \rightarrow |b,s\rangle $ and by $\Delta _{ss}$ from $|c,s\rangle
\rightarrow |d,s,s\rangle $. $\hat{E}_{s,\pm }(z,t)$\ also drive the
transitions from $|c,s\rangle \rightarrow |d,s,\bar{s}\rangle $\ with
detuning $\Delta _{s\bar{s}}$. Here $\bar{s}=\uparrow ,\downarrow $\ and $%
\bar{s}\neq s$. Finally, the applied classical control beams with Rabi
frequencies $\Omega _{s,\pm }(t)$ couple to both atoms and drive the
transitions $|b,s\rangle \rightarrow |c,s\rangle $.


The evolution of the slowly-varying quantum operators $\hat{E}_{s,\pm }$ are
given by four Maxwell-Bloch (MB) equations
\begin{equation}
\left( \partial _{t}+\nu \partial _{z}\right) \hat{E}_{s,\pm }=\mathrm{i}%
\sqrt{2\pi }n_{z}^{\mathnormal{s}}g\left( \sigma _{a;b,s,\pm }+\sigma
_{c,s;d,s,s,\pm }+\sigma _{c,\bar{s};d,s,\bar{s},\pm }\right)  \label{MB}
\end{equation}%
with four levels of the $s$-th atoms denoted as $|a,s\rangle $, $|b,s\rangle
$\ , $|c,s\rangle $\ and $|d,s,s\rangle $. When writing down the MB
equations (\ref{MB}), we have introduced the slowly-varying collective
operators
\begin{equation}
\sigma _{\mu ;\nu }=\sigma _{\mu ;\nu ,+}(z,t)\mathrm{e}^{\mathrm{i}k_{%
\mathrm{Q}\mathnormal{,s}}z}+\sigma _{\mu ;\nu ,-}(z,t)\mathrm{e}^{-\mathrm{i%
}k_{\mathrm{Q}\mathnormal{,s}}z},
\end{equation}%
and $\nu $ is the velocity of quantum fields in an empty waveguide \footnote{%
We would like to highlight here the relative simplicity of the above
evolution equation compared to the one we considered in \cite{SCS}\ where
extra phase terms have to be involved due to the existence of
two-atomic-species different frequencies on the incident quantum fields.}.
In the derivation of equations of motion we assume the Rabi frequencies of
the control fields to be slowly varied. The slow-light polariton operators
are defined as
\begin{equation}
\Psi _{s,\pm }=\cos \theta _{s}\hat{E}_{s,\pm }-\sin \theta _{s}\sqrt{2\pi
n_{z}^{\mathnormal{s}}}\sigma _{c,s;a},
\end{equation}%
where $\tan \theta _{s}=g\sqrt{2\pi n_{z}^{\mathnormal{s}}}/\Omega _{s}$.
For stationary polaritons we have assumed that the amplitudes of the
counterpropagating classical fields are equal, \textit{i.e.}, $\Omega
_{s,\pm }\equiv \Omega _{s}$. In the limit when the excitations are mostly
in spin-wave form, \textit{i.e.}, $\sin \theta _{s}\simeq 1$, and since $%
\sigma _{c,s;a}=-g\hat{E}_{s,\pm }/\Omega _{s}$, the polariton operators
become
\begin{equation}
\Psi _{s,\pm }=\sqrt{2\pi n_{z}^{\mathnormal{s}}}\frac{g}{\Omega _{s}}\hat{E}%
_{s,\pm }.
\end{equation}%
Setting $\Psi _{s}=(\Psi _{s,+}+\Psi _{s,-})/2$ and $A_{s}=(\Psi _{s,+}-\Psi
_{s,-})/2$ as the symmetric and antisymmetric combinations of the two
polaritons, we derive the equations of motion for the polariton combinations
$\Psi _{s},\;A_{s}$:
\begin{eqnarray}
\partial _{t}\Psi _{s} &+&\nu \partial _{z}A_{s}=-\pi \tan ^{2}\theta
_{s}\partial _{t}\Psi _{s}-\mathrm{i}\frac{2\pi g^{2}}{\Delta _{ss}}\left(
2\Psi _{s}^{\dagger }\Psi _{s}+A_{s}^{\dagger }A_{s}\right) \Psi _{s}
\nonumber \\
&&-\mathrm{i}\frac{2\pi g^{2}}{\Delta _{s\bar{s}}}\left( 2\Psi _{\bar{s}%
}^{\dagger }\Psi _{\bar{s}}+A_{\bar{s}}^{\dagger }A_{\bar{s}}\right) \Psi
_{s}+\mathrm{noise},  \nonumber \\
\partial _{t}A_{s} &+&\nu \partial _{z}\Psi _{s}=-\mathrm{i}\frac{2\pi g^{2}%
}{\Delta _{s}}n_{z}^{\mathnormal{s}}A_{s}-\frac{2\pi g^{2}}{\Delta _{ss}}%
\Psi _{s}^{\dagger }\Psi _{s}A_{s}+\mathrm{noise}.  \label{eom}
\end{eqnarray}

The noise terms in Eq. (\ref{eom}) account for the dissipative processes
that take place during the evolution. Fortunately, for the dark-state
polaritons under consideration, as long as the spontaneous emission rates $%
\Gamma $ from the states $|c,s\rangle $ and $|d,s,s^{\prime }\rangle $ are
much less than the detunings $|\Delta _{ss^{\prime }}|$, the losses in the
timescales of interest are not significant and thus can be neglected as
discussed in \cite{DPS1,DPS2,DPS3,Chang,SCS}. Assuming optical depth of a
few thousand and a large ratio between the density of atoms to the density
of photons $n_{z}^{s}/n_{\mathrm{ph}}^{s}\sim 10^{4}$, the antisymmetric
combinations $A_{\uparrow }$ and $A_{\downarrow }$ can be adiabatically
eliminated from the equations of motion for the polaritons and moreover, the
nonlinear terms like $\Psi _{s}^{\dagger }\Psi _{s}A_{s}$\ and $%
A_{s}^{\dagger }A_{s}\Psi _{s}$\ are negligible. In this regime, Eq. (\ref%
{eom}) simplifies to a nonlinear Schr\"{o}dinger Eq. (\ref{NLE}) for
polaritons which reads: 
\begin{equation}
i\partial _{t}\Psi _{s}=\frac{2\Delta _{s}\nu _{s}}{\Gamma _{\mathrm{1D}%
}^{s}n_{z}^{\mathnormal{s}}}\partial _{z}^{2}\Psi _{s}+\frac{\Gamma _{%
\mathrm{1D}}^{s}\nu _{s}}{\Delta _{ss}}\Psi _{s}^{\dagger }\Psi _{s}\Psi
_{s}+\frac{\Gamma _{\mathrm{1D}}^{s}\nu _{s}}{\Delta _{s\bar{s}}}\Psi _{\bar{%
s}}^{\dagger }\Psi _{\bar{s}}\Psi _{s},  \label{NLE}
\end{equation}%
which is related to an effective two-component Lieb-Liniger model of
polaritons%
\begin{equation}
H=\int dz\sum_{s}\left[ \frac{1}{2m_{s}}\partial _{z}\Psi _{s}^{\dagger
}(z)\partial _{z}\Psi _{s}(z)+\frac{\chi _{s}}{2}\rho _{s}^{2}(z)\right]
+\int dz\chi _{\uparrow \downarrow }\rho _{\uparrow }(z)\rho _{\downarrow
}(z).  \label{mdp}
\end{equation}%
Here $m_{s}=-\Gamma _{\mathrm{1D}}^{s}n_{z}^{\mathnormal{s}}/(4\Delta
_{s}\nu _{s})$ is the effective mass for $s$-th polaritons with $\Gamma _{%
\mathrm{1D}}^{s}=4\pi g^{2}/\nu $ the spontaneous emission rate of a single
atom into the waveguide modes and $\nu _{s}=\nu \Omega _{s}^{2}/\left( \pi
g^{2}n_{z}^{\mathnormal{s}}\right) $ the group velocity of the propagating
polaritons. The intraspecies repulsion is characterized by $\chi _{s}=\Gamma
_{\mathrm{1D}}^{s}\nu _{s}/\Delta _{ss}$ and the interspecies repulsion by $%
\chi _{\uparrow \downarrow }=\sum_{s=\uparrow ,\downarrow }\Gamma _{\mathrm{%
1D}}^{s}\nu _{s}/\Delta _{s\bar{s}}$.

To reach the spin-charge separation regime, we employ the mapping to a
Luttinger liquid model described in the previous section. For this to be
possible, we should first of all check the tunability of the relevant
parameters to the repulsive regime. This in our case (see Eq. (\ref{NLE}))
implies tuning $\Delta _{s}\Delta _{ss}<0$ which leads to $m_{s}\chi _{s}>0$%
. Similarly $\Delta _{\uparrow }\Delta _{\downarrow }>0$ forces the
effective masses $m_{\uparrow }m_{\downarrow }>0$ and $\Delta _{\uparrow
\uparrow }\Delta _{\downarrow \downarrow }>0$ tunes $\chi _{\uparrow }\chi
_{\downarrow }>0$. These conditions are satisfied by tuning the lasers such
that: $\Delta _{\uparrow }$ and$\ \Delta _{\downarrow }$\ are negative
(positive) while at the same time $\Delta _{\uparrow \uparrow }$\ and $%
\Delta _{\downarrow \downarrow }$ are positive (negative).

Apart from the repulsive interaction regime, the separation condition Eq. (%
\ref{con}) needs to be satisfied as well, which in our case means setting:
\begin{equation}
\chi _{\uparrow }=\chi _{\downarrow },\frac{\rho _{0,\uparrow }}{m_{\uparrow
}}=\frac{\rho _{0,\downarrow }}{m_{\downarrow }},  \label{sepcon}
\end{equation}%
where the polariton density $\rho _{0,s}$ equals to the photon density $n_{%
\mathrm{ph}}^{s}$. The effective charge and spin densities are the sum and
difference of the two-species polaritonic densities, which read as%
\begin{equation}
\rho _{\mathrm{charge}}=\rho _{\uparrow }+\rho _{\downarrow },\rho _{\mathrm{%
spin}}=\rho _{\uparrow }-\rho _{\downarrow }
\end{equation}%
with $\rho _{s}=\left( \rho _{0,s}-\frac{1}{\pi }\nabla \phi _{s}\right)
\sum_{m}\exp \left[ i2m\left( \pi \rho _{0,s}z-\phi _{s}\right) \right] $.\
Keeping only the lowest components with $m=0,\pm 1$, the charge and spin
density operators in the bosonic language can be represented as
\begin{eqnarray}
\rho _{\mathrm{charge}} &=&\rho _{0}-\frac{\sqrt{2}}{\pi }\partial _{z}\phi
_{\mathrm{charge}}+2\rho _{0}\cos \left[ 2k_{\mathrm{F}}z-\sqrt{2}\phi _{%
\mathrm{charge}}\right] \cos \sqrt{2}\phi _{\mathrm{spin}},  \label{dcharge}
\\
\rho _{\mathrm{spin}} &=&-\frac{\sqrt{2}}{\pi }\partial _{z}\phi _{\mathrm{%
spin}}+2\rho _{0}\sin \left[ 2k_{\mathrm{F}}z-\sqrt{2}\phi _{\mathrm{charge}}%
\right] \sin \sqrt{2}\phi _{\mathrm{spin}}.
\end{eqnarray}%
Here the first term in $\rho _{\mathrm{charge}}$ is the average density $%
\rho _{0}=\rho _{0,\uparrow }+\rho _{0,\downarrow }$. In our two-species
photonic system, we set $\rho _{0,\uparrow }=\rho _{0,\downarrow }=\frac{1}{2%
}\rho _{0}$ for each polarization component. The second gradient term in $%
\rho _{\mathrm{charge}}$ and $\rho _{\mathrm{spin}}$ are the density
oscillations with zero momentum. The third term in $\rho _{\mathrm{charge}}$%
\ and second term in $\rho _{\mathrm{spin}}$\ are the density fluctuations
of the $2k_{\mathrm{F}}$ components \cite{Iucci}.\ We label $\gamma _{s}$ as
the ratio of the interaction to the kinetic energies for each polariton
species $\gamma _{s}=m_{s}\chi _{s}/\rho _{0,s}$. Combining the two
separation conditions in Eq. (\ref{sepcon})\ together, one gets $\gamma
_{\uparrow }=\gamma _{\downarrow }$. For $\chi =\chi _{s}$ and $\gamma
=\gamma _{s}$, the velocities and Luttinger parameters can be expressed as $%
u=\chi /\sqrt{\gamma }$ and $K=\pi /\sqrt{\gamma }$. As also demonstrated
for a similar system albeit with one quantum field \cite{Chang}, $\gamma $
here can also be tuned from zero to finite to extremely large, corresponding
to non-, weak- and strong-correlated regimes, which implies a wide tunable
range for $u$\ and $K$.

\subsection{Probing of the photonic spinons and holons}


\begin{figure}[tbp]
\centering
\includegraphics[width = 1\linewidth]{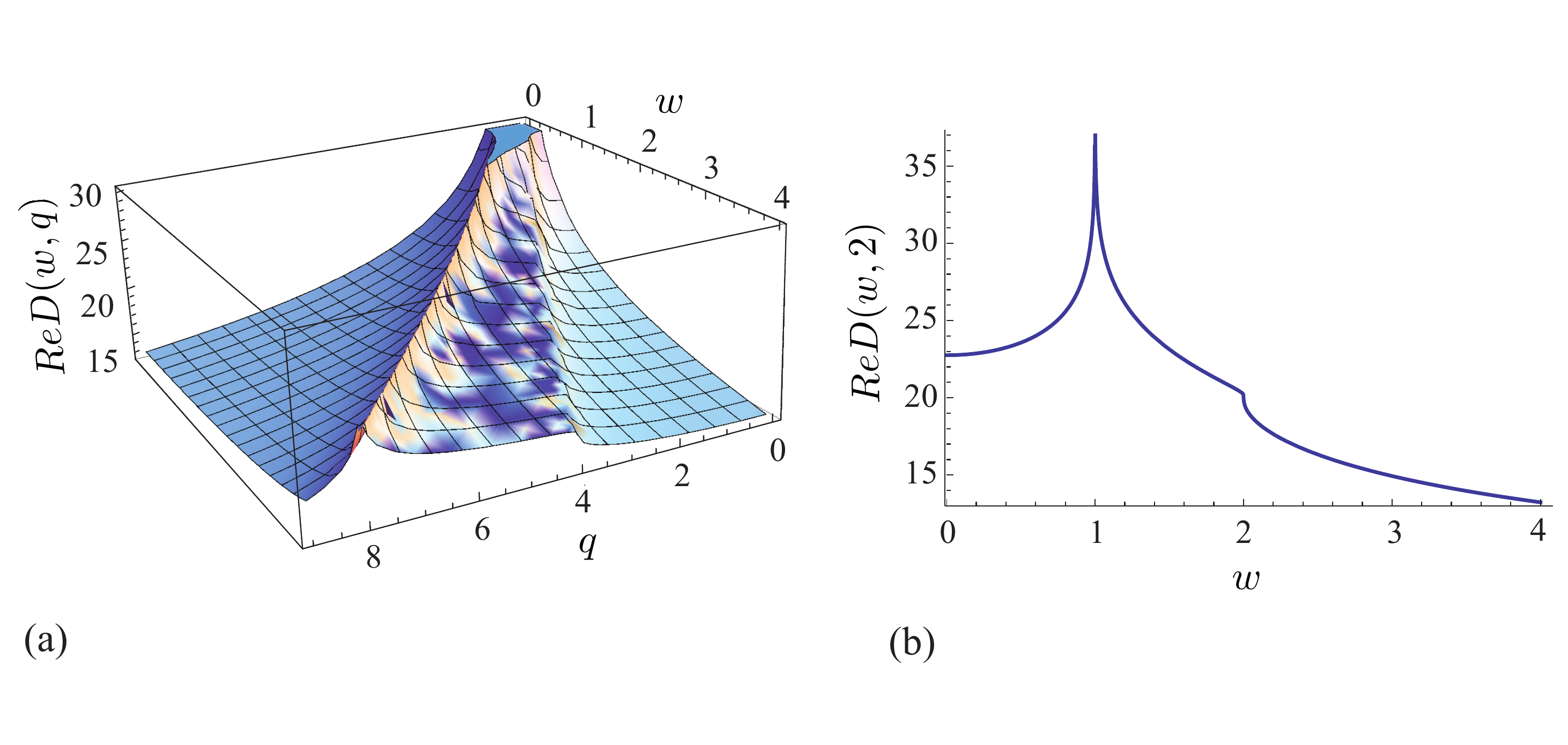}
\caption{The Fourier transform of density-density correlations for
our polaritonic system. It exhibits the characteristic splitting
corresponding to the two different propagation velocities for the
photonic spinons and holons, as calculated in \protect\cite{Iucci}.
The effective spin and charge velocities are $u_{\mathrm{charge}}=1$
and $u_{\mathrm{spin}}=0.5$ (normalized by
$2\protect\sqrt{\frac{2}{5}}u$), and the Luttinger parameters are
$K_{\mathrm{charge}}=0.55$ and $K_{\mathrm{spin}}=1.1$. They
translate in our case at optical depths of $OD=2000$, $10$ photons
in each pulse initially, and single-atom co-operativity for each
atomic species of $0.4$. In (b) a cut of the 3D plot at
quasimomentum $q=2$ is plotted to show the distinct two peaks
corresponding to the spin and charge velocities. The cross
density-density correlations are experimentally reconstructed via
typical optical measurements on the correlated photon states as they
exit the fiber.} \label{fc}
\end{figure}

The typical detection of spin-charge separation can occur through
dynamically probing the time evolution of a single excitation as in
cold-atom proposals \cite{Recati1,Recati2,Recati3,Recati4}, or by measuring
the corresponding single-particle spectral function as in condensed matter
experiments \cite%
{Girardeau1,Girardeau2,Girardeau3,Girardeau4,SC_exp_old_77,SC_exp_old_402,SC_exp_old_418, SCC4_308, SCC4_2, SCC4_325}%
. In our case, we propose to extract the charge and spin velocities by
measuring the Fourier transform of density-density correlations $D(\omega ,q)
$ for energy $\omega $ and momentum $q$. As derived for the two-component
system in \cite{Iucci}, for the $2k_{\mathrm{F}}$\ component of density
operator $\rho _{2k_{\mathrm{F}}}=2\rho _{0}\cos \left( 2k_{\mathrm{F}}z-%
\sqrt{2}\phi _{\mathrm{charge}}\right) \cos \sqrt{2}\phi _{\mathrm{spin}}$
(the last term in Eq. (\ref{dcharge})), the Fourier transform $D(\omega ,q)$%
\ of the density-density operator $\left\langle T_{\tau }\rho _{2k_{\mathrm{F%
}}}\left( z,\tau \right) \rho _{2k_{\mathrm{F}}}\left( 0,0\right)
\right\rangle $ is given by%
\begin{eqnarray}
D(\omega ,q) &=&-\frac{4\pi \rho _{0}^{2}\left( \frac{\alpha }{2}\right)
^{K_{\mathrm{charge}}+K_{\mathrm{spin}}}\Gamma \left( 1-K_{\mathrm{charge}%
}/2-K_{\mathrm{spin}}/2\right) }{\Gamma \left( K_{\mathrm{charge}}/2+K_{%
\mathrm{spin}}/2\right) }  \nonumber \\
&&\times \left\vert \omega ^{2}-u_{\mathrm{spin}}^{2}q^{2}\right\vert ^{K_{%
\mathrm{charge}}/2+K_{\mathrm{spin}}/2-1}u_{\mathrm{spin}}^{1-K_{\mathrm{%
charge}}-K_{\mathrm{spin}}}  \nonumber \\
&&\times \exp \left[ -i\pi \left( \frac{K_{\mathrm{charge}}+K_{\mathrm{spin}}%
}{2}-1\right) \Theta \left( \omega ^{2}-u_{\mathrm{spin}}^{2}q^{2}\right) %
\right]   \nonumber \\
&&\times F_{1}\big(\frac{K_{\mathrm{charge}}}{2},\frac{K_{\mathrm{charge}%
}+K_{\mathrm{spin}}-1}{2},1-\frac{K_{\mathrm{charge}}+K_{\mathrm{spin}}}{2},
\nonumber \\
&&\frac{K_{\mathrm{charge}}+K_{\mathrm{spin}}}{2};1-\frac{u_{\mathrm{charge}%
}^{2}}{u_{\mathrm{spin}}^{2}},1-\frac{\omega ^{2}-u_{\mathrm{charge}%
}^{2}q^{2}}{\omega ^{2}-u_{\mathrm{spin}}^{2}q^{2}}\big),
\end{eqnarray}%
where $\Gamma $ is the gamma function, $\Theta $ is the step function, $F_{1}
$ is the Appell's hypergeometric function, and $\alpha $ is a short-distance
cutoff. $D(\omega ,q)$ depends on the velocity $\omega /q$ and should
exhibit two peaks centered around $u_{\mathrm{charge}}q$ and $u_{\mathrm{spin%
}}q$ \cite{Girardeau1,Girardeau2,Girardeau3,Girardeau4}. In our
photonic system, probing of the spinon and holon branches can be
done by measuring the correlation functions of densities of the
fields as they exit, for a specific quasi-momenta $q$. For a clear
distinction between the two effective spin and charge peaks, we
should set our optical detectors around $q=2\pi /z_{0}$, \textit{i.e.}, $z_{0}$ apart ($%
z_{0}$ here corresponds to roughly the length of the fiber). To give an
illustration of the expected behavior, $D(\omega ,q)$\ in the unit of $\rho
_{0}^{2}\alpha $\ is plotted in figure \ref{fc} with the intra- and
interspecies repulsion ratio $\chi _{\uparrow \downarrow }/\chi =0.6$, which
in turn tunes the charge and spin velocities to
\begin{equation}
u_{\mathrm{charge}}=2u_{\mathrm{spin}}=u\sqrt{1+\chi _{\uparrow \downarrow
}/\chi }=2\sqrt{\frac{2}{5}}u,
\end{equation}%
and Luttinger parameters $K_{\mathrm{charge}}=\frac{1}{2}K_{\mathrm{spin}}=%
\frac{1}{2}\sqrt{\frac{5}{2}}K$. We choose $u_{\mathrm{charge}}=1$, $u_{%
\mathrm{spin}}=0.5$ and $K_{\mathrm{charge}}=0.55$, $K_{\mathrm{spin}}=1.1$\
via tuning $u$\ and $K$\ wihch require $\gamma \sim 20$. This as shown in
\cite{Chang} is achieved at optical depths $OD=2000$ and roughly $N_{%
\mathnormal{ph}}^{\uparrow ,\downarrow }=10$ photons initially in each pulse
and single-atom co-operativity of $\eta =0.4$ \footnote{%
Co-operativity here is the ratio of spontaneous emission into the waveguide
to total spontaneous emission.}. These values are for the moment out of the
current experimental range where optical depths of a few hundred have been
achieved, but should not be out of the question in the near to mid-term
future \cite{Nayak,Vetsch,Bajcsy2011}. In calculating the optical
interaction parameters appearing in the Hamiltonian Eq. (20), we haven take
into account both the linear and nonlinear loss mechanisms as layed out in
\cite{Chang}.

\section{Conclusion}

We have described in detail a strongly correlated photonic scheme to
simulate a purely fermionic effect, spin-charge separation. In more detail,
we have shown that polarized photons interacting with a cold atomic ensemble
can be made to obey two-component Lieb-Liniger physics and even behave as a
quantum Luttinger liquid. The relevant interactions exhibit the necessary
tunability for steering the photons to the effective spin-charge separation
regime. Efficient observations of the characteristic features of the
separation using standard quantum optical methods should be feasible based
on correlations measurements of the outgoing photons which here carry
opposite polarizations. The current proposal is different from a similar
scheme proposed earlier by some of us, where two species of atoms were
coupled to two quantum fields of two different frequencies but of same
polarization \cite{SCS}. Here a single species of atoms is shown to suffice
in order to induce the required intra- and interspecies interactions, which
combined with the easier detection of the polarized output states makes this
approach more feasible.

\section{Acknowledgments}

We would like to acknowledge financial support by the National Research
Foundation \& Ministry of Education, Singapore.


\section*{References}

\begin{thebibliography}{99}
\bibitem{Lieb} Lieb E H and Liniger W 1963 \textit{Phys. Rev.} \textbf{130}
1605

\bibitem{Girardeau1} Giamarchi T 2004 \textit{Quantum Physics in One
Dimension} (Oxford University Press, Oxford)

\bibitem{Girardeau2} Girardeau M 1960 \textit{J. Math. Phys.} \textbf{1}
516-23

\bibitem{Girardeau3} Girardeau M 1965 \textit{Phys. Rev.} \textbf{139} B500

\bibitem{Girardeau4} Paredes B \textit{et al} 2004 \textit{Nature} \textbf{%
429} 277-81

\bibitem{SC_exp_old_77} Kim C \textit{et al} 1996 \textit{Phys. Rev. Lett.}
\textbf{77} 4054-7

\bibitem{SC_exp_old_402} Segovia P, Purdie D, Hengsberger M and Baer Y 1999
\textit{Nature} \textbf{402} 504-7

\bibitem{SC_exp_old_418} Lorenz T, Hofmann M, Gr\"{u}ninger M, Freimuth A,
Uhrig G S, Dumm M and Dressel M 2002 \textit{Nature} \textbf{418} 614-7

\bibitem{SCC4_308} Auslaender O M \textit{et al} 2005 \textit{Science}
\textbf{308} 88-92

\bibitem{SCC4_2} Kim B J \textit{et al} 2006 \textit{Nat. Physics} \textbf{2}
397-401

\bibitem{SCC4_325} Jompol Y \textit{et al} 2009 \textit{Science} \textbf{325}
597-601


\bibitem{Recati1} Recati A, Fedichev P O, Zwerger W and Zoller P 2003
\textit{Phys. Rev. Lett.} \textbf{90} 020401

\bibitem{Recati2} Kecke L, Grabert H and Hausler W 2005 \textit{Phys. Rev.
Lett.} \textbf{94} 176802

\bibitem{Recati3} Kollath C, Schollw\"{o}ck U and Zwerger W 2005 \textit{%
Phys. Rev. Lett.} \textbf{95} 176401

\bibitem{Recati4} Kleine A, Kollath C, McCulloch I P, Giamarchi T and Schollw%
\"{o}ck U 2008 \textit{Phys. Rev. A} \textbf{77} 013607

\bibitem{SIPS1} Angelakis D G, Santos M F, Yannopapas V and Ekert A 2007
\textit{Phys. Lett. A} \textbf{362} 377-80

\bibitem{SIPS21} Angelakis D G, Santos M F and Bose S 2007 \textit{Phys.
Rev. A} \textbf{76} 031805(R)

\bibitem{SIPS22} Hartmann M J, Brand\~ao F G S L and Plenio M B 2006 \textit{%
Nat. Phys.} \textbf{2} 849-55

\bibitem{SIPS23} Greentree A D, Tahan C, Cole J H and Hollenberg L C L 2006
\textit{Nat. Phys.} \textbf{2} 856-61

\bibitem{SIPSother1} Rossini D and Fazio R 2007 \textit{Phys. Rev. Lett.}
\textbf{99} 186401

\bibitem{SIPSother2} Na N, Utsunomiya S, Tian L and Yamamoto Y 2008 \textit{%
Phys. Rev. A} \textbf{77} 031803(R)

\bibitem{SIPSother3} Aichhorn M, Hohenadler M, Tahan C and Littlewood P B
2008 \textit{Phys. Rev. Lett.} \textbf{100} 216401

\bibitem{SIPSother4} Gerace D, T\"{u}reci H E, Imamoglu A, Giovannetti V and
Fazio R 2009 \textit{Nat. Phys.} \textbf{5} 281-4

\bibitem{SIPSother5} Carusotto I \textit{et al} 2009 \textit{Phys. Rev. Lett.%
} \textbf{103} 033601

\bibitem{SIPSother6} Angelakis D G, Bose S and Mancini S 2009 \textit{Eur.
Phys. Lett.} \textbf{85} 20007

\bibitem{DPS1} Fleischhauer M and Lukin M D 2000 \textit{Phys. Rev. Lett.}
\textbf{84} 5094-7

\bibitem{DPS2} Bajcsy M, Zibrov A S and Lukin M D 2003 \textit{Nature}
\textbf{426} 638-41

\bibitem{DPS3} Bajcsy M \textit{et al} 2009 \textit{Phys. Rev. Lett.}
\textbf{102} 203902

%

\bibitem{Chang} Chang D E \textit{et al} 2008 \textit{Nat. Phys.} \textbf{4}
884-9

\bibitem{SCS} Angelakis D G, Huo M-X, Kyoseva E and Kwek L C 2011 \textit{%
Phys. Rev. Lett.} \textbf{106} 153601


\bibitem{Nayak} Nayak K P \textit{et al} 2007 \textit{Opt. Express} \textbf{%
15} 5431-8

\bibitem{Vetsch} Vetsch E, Reitz D, Sague G, Schmidt R, Dawkins S T, and
Rauschenbeutel A 2010 \textit{Phys. Rev. Lett.} \textbf{104} 203603

\bibitem{book_Menzel1} Ghosh S, Sharping J E, Ouzounov D G and Gaeta A L
2005 \textit{Phys. Rev. Lett.} \textbf{94} 093902

\bibitem{book_Menzel3} Takekoshi T and Knize R J 2007 \textit{Phys. Rev.
Lett.} \textbf{98} 210404

\bibitem{book_Menzel4} Christensen C A \textit{et al} 2008 \textit{Phys.
Rev. A} \textbf{78} 033429

\bibitem{book_Menzel5} Vorrath S, M\"{o}ller S A, Windpassinger P, Bongs K
and Sengstock K 2010 \textit{New J. Phys.} \textbf{12} 123015

\bibitem{Iucci} Iucci A, Fiete G A and Giamarch T 2007 \textit{Phys. Rev. B}
\textbf{75} 205116

\bibitem{Bajcsy2011} Bajcsy M \textit{et al} 2011 \textit{Phys. Rev. A}
\textbf{83} 063830
\end{thebibliography}
\end{document}